\documentclass[12pt,english,sort&compress]{article}
\usepackage[T1]{fontenc}
\usepackage[latin9]{inputenc}
\usepackage{fancyhdr}
\usepackage{babel}
\usepackage{booktabs}
\usepackage{amsbsy}
\usepackage{graphicx}
\usepackage[a4paper]{geometry}
\usepackage{setspace}
\usepackage{esint}
\usepackage[numbers]{natbib}
\usepackage[bookmarks=true,bookmarksnumbered=false,bookmarksopen=false,
 breaklinks=false,pdfborder={0 0 0},pdfborderstyle={},backref=false,colorlinks=false]
 {hyperref}
\hypersetup{
 pdfauthor={Noy Cohen}}

\makeatletter

\providecommand{\tabularnewline}{\\}

\usepackage{times}

\date{}

\renewcommand{\@openbib@code}{\setlength{\itemsep}{-1pt}}

\renewcommand{\subsectionmark}[1]{}

\usepackage[compact]{titlesec}
\titleformat{\section}{\LARGE \bfseries}{\thesection}{1em}{}
\titleformat{\subsection}{\large \bfseries}{\thesubsection}{1em}{}

\makeatother

\begin{document}
\global\long\def\d{\mathrm{d}}%
\global\long\def\boltzmann{k}%

\global\long\def\chains{N_{0}}%
\global\long\def\reflen{L_{0}}%
\global\long\def\refdiam{D_{0}}%
\global\long\def\curlen{l}%
\global\long\def\curdiam{d}%

\global\long\def\defgradT{\mathbf{F}}%
\global\long\def\xh{\hat{\mathbf{x}}}%
\global\long\def\yh{\hat{\mathbf{y}}}%
\global\long\def\zh{\hat{\mathbf{z}}}%

\global\long\def\stress{\sigma}%
\global\long\def\stretch{\lambda}%
\global\long\def\E{E}%

\global\long\def\energy{\psi}%
\global\long\def\energyb{\psi^{\left(b\right)}}%
\global\long\def\energyn{\psi^{\left(n\right)}}%

\global\long\def\defgradTeb{\mathbf{F}_{e}^{\left(b\right)}}%
\global\long\def\defgradTpb{\mathbf{F}_{p}^{\left(b\right)}}%
\global\long\def\defgradTen{\mathbf{F}_{e}^{\left(n\right)}}%

\global\long\def\mandel{M}%
\global\long\def\yieldstress{\stress_{y}}%
\global\long\def\yieldsurface{f}%
\global\long\def\yieldstressf{\mu_{y}}%

\global\long\def\stretcheb{\lambda_{e}^{\left(b\right)}}%
\global\long\def\stretchpb{\lambda_{p}^{\left(b\right)}}%
\global\long\def\stretchen{\lambda_{e}^{\left(n\right)}}%

\global\long\def\stretchul{\lambda_{ul}}%
\global\long\def\stretchpul{\lambda_{ul}^{\left(p\right)}}%
\global\long\def\stretchr{\lambda^{\left(r\right)}}%
\global\long\def\stretchl{\lambda^{\left(l\right)}}%

\global\long\def\stressb{\sigma^{\left(b\right)}}%
\global\long\def\stressn{\sigma^{\left(n\right)}}%

\global\long\def\yieldc{h}%
\global\long\def\plastic{\gamma}%

\title{\vspace{-80pt}Plasticity, hysteresis, and recovery mechanisms in
spider silk fibers}
\author{Renata Oliv{\'e} $^{\footnotesize{a}}$, Jos{\'e} P{\'e}rez-Riguero $^{\footnotesize{b,c}}$,
and Noy Cohen $^{\footnotesize{a,}} \footnote{e-mail address: noyco@technion.ac.il}\,\,$
\\
{\footnotesize \parbox[t]{\linewidth}{\centering 
{\normalsize $^{a}$} Department of Materials Science and Engineering, Technion  - Israel Institute of Technology, Haifa 3200003, Israel
}}\\
{\footnotesize \parbox[t]{\linewidth}{\centering 
{\normalsize $^{b}$} Center for Biomedical Technology, Universidad Polit{\'e}cnica de Madrid, 28223 Pozuelo de Alarc{\'e}n (Madrid), Spain
}}\\
{\footnotesize \parbox[t]{\linewidth}{\centering 
{\normalsize $^{c}$} Departamento de Ciencia de Materiales, ETSI Caminos, Canales y Puertos, Universidad Polit{\'e}cnica de Madrid, 28040 Madrid, Spain
}}
\vspace{-1.5em}}
\maketitle
\begin{abstract}
Spider silk is a remarkable biomaterial with exceptional stiffness,
strength, and toughness stemming from a unique microstructure. While
recent studies show that silk fibers exhibit plasticity, hysteresis,
and recovery under cyclic loading, the underlying microstructural
mechanisms are not yet fully understood. In this work, we propose
a mechanism explaining the loading-unloading-relaxation response through
microstructural evolution: initial loading distorts intermolecular
bonds, resulting in a linear elastic regime. Upon reaching the yield
stress, these bonds dissociate and the external load is transferred
to the polypeptide chains, which deform entropically to allow large
deformations. Unloading is driven by entropic shortening until a traction
free state with residual stretch is achieved. Subsequently, the fiber
recovers as chains reorganize and bonds reform, locking the microstructure
into a new stable equilibrium that increases stiffness in subsequent
cycles. Following these mechanisms, we develop a microscopically motivated,
energy-based model that captures the macroscopic response of silk
fibers under cyclic loading. The response is decoupled into two parallel
networks: (1) an elasto-plastic network of inter- and intramolecular
bonds governing the initial stiffness and yield stress, and (2) an
elastic network of entropic chains that enable large deformations.
The model is validated against experimental data from \emph{Argiope
bruennichi} dragline silk. The findings from this work are three-fold:
(1) explaining the mechanisms that govern hysteresis and recovery
and linking them to microstructural evolution; (2) quantifying the
recovery process of the fiber, which restores and enhances mechanical
properties; and (3) establishing a predictive foundation for engineering
synthetic fibers with customized properties.
\end{abstract}


\section{Introduction}

Spider silk fibers are renowned for their exceptional mechanical performance,
which includes high stiffness, high tensile strength, and remarkable
toughness \citep{elic&etal11JMBBM,port&voll08SM,du&etal06BJ,voll&etal96PRSB,gosl&etal84Nature,roem&sche08Prion,yarg&etal18NRM}.
In addition, exposure of these silk fibers to high humidity results
in supercontraction, or a shortening of up to 60\% in length \citep{work&moro82TRJ,cohe&etal21BMM,guin&etal05JEB,plaz&etal06JPS,fazi&etal22JMPS,PerezRigueiro2003,cohe23JMPS,Fazio2023},
and twist \citep{xu&etal14SM,liu&etal19SA,cohe&eise22PRL}. Owing
to this unique combination of strength and extensibility, spider silk
has attracted significant attention for applications across diverse
fields, ranging from tissue engineering and regenerative medicine
\citep{Esser2021,Koeck2024,Zhang2021} to high-performance biocompatible
composite materials \citep{Archana2018,lefe&auge16IMR,Qin2015}. Recent
works also proposed generative modeling and design methods through
the spider silk protein sequences to enhance the mechanical properties
and response of silk-based materials \citep{Lu2024,Lu2025}.

The fiber properties stem from a unique semi-crystalline microstructure
comprising crystalline domains, which act as permanent cross-linkers,
that are embedded in an amorphous protein matrix \citep{kete&etal10NM,Nova2010}.
More specifically, the crystalline domains are made of stacked poly(alanine)
$\beta$-sheets that interconnect glycine-rich polypeptide chain segments.
In the dragline spider silk fiber, the chains are typically highly
extended and aligned along the fiber direction \citep{Olive2024,madr&etal16SR,plaz&etal12SM}.
This configuration, which is not energetically favorable, is maintained
by a series of intermolecular hydrogen bonds that restrict chain mobility.
Additionally, experimental evidence shows that the polypeptide chains
contain intramolecular $\beta$-sheets, which store a compact \textquotedbl hidden
length\textquotedbl{} that can unravel and increase the contour length
of the chains once a sufficient force is applied \citep{beek&etal02PNAS,du&etal11AFM,du&etal06BJ,Olive2024}.
This effect contributes to the extensibility of the fiber. 

The origin of the mechanical response of the fibers has been widely
explored both experimentally and theoretically. Many works showed
that the response of silk fibers depends on many factors, including
the spinning conditions (i.e. naturally spun or forcibly silked),
the spider species, and the reeling speed \citep{guin&etal05JEB,Madsen1999,Liu2005,elic&etal05JOM,Olive2025}.
Broadly, these govern the microstructure of the fiber and the network
configuration and are responsible for the formation of crystalline
domains, intercrystallite distance, chain alignment, initial chain
extensions, and the number of intermolecular hydrogen bonds \citep{term94MM,Cohen2025,elic&etal09BMM,Chen2006,Young2021,Olive2025}.
Recent works also showed that high strain rates lead to an increase
in toughness \citep{Hudspeth2012,yaza&etal20CM}. In the context of
wetting-drying cycles, \citet{blac&etal09JEB} investigated the response
of fibers and showed a reversible uptake of water, as well as reduced
stiffness and yield stress post-supercontraction. Not only the strain
rate, but also the hydration rate plays was found to play a role in
supercontraction \citep{agna&etal09Zoology}. The mechanical response
under cyclic loading of braided spider silk sutures and \emph{Bombyx
mori }silk fibroin were also studied \citep{Hennecke2013,Patel2022}. 

One of the less investigated key features of the spider silk fiber
is its plasticity, hysteresis, and recovery under cyclic loading.
The response under this loading condition is crucial for the biological
function of the silk and the understanding of the mechanisms that
govern plasticity, hysteresis, and recovery can lead to tools that
enable one to manipulate and control the behavior of the fiber over
time. Recent experimental \citep{Jiang2023,Chen2006,veho&etal07BJ,Yu2015}
and theoretical \citep{tomm&etal10BJ,Jiang2020a,Jiang2020,term94MM,veho&etal07BJ,Patil2022}
works investigated the response of spider silk fibers under uniaxial
cyclic loading. These focused on the ability of the fiber to dissipate
energy by examining the hysteresis loop in the measured stress-strain
curves of different cycles. Upon loading, fibers exhibit a glassy
behavior - the initial response is linear up to a yield stress, followed
by a plateau and a strain stiffening effect. During this process,
the fiber exhibits elasto-plastic mechanisms. Once unloaded, the fiber
``relaxes'' with a curve that is typical to rubbery polymers, with
a traction free stress that is characterized by a plastic deformation
(i.e. a residual strain). Next, the fiber is allowed to relax. During
this process, the fiber can exhibit a slight decrease in the plastic
strain due to the relaxation of the fiber. Additional loading cycles
typically show the same characteristic response. 

This work aims to develop a fundamental microscopically motivated
energy-based model that sheds light on the underlying mechanisms that
govern the response of spider silk fibers under cyclic loading. Specifically,
we focus on the evolution of the microstructural quantities that govern
plasticity, hysteresis, and recovery and link these to the macroscopic
performance of the fiber. To this end, we propose to decouple the
response of the fiber into two mechanisms that act in parallel: (1)
an elasto-plastic network made of intermolecular and intramolecular
hydrogen bonds that dissociate in response to tension and govern the
initial (linear) response of the fiber and (2) an elastic network
comprising entropic polypeptide chains that provide the elasticity
of the network under large deformations. During the initial loading,
the first network is dominant. Once the yield stress is reached, the
bonds dissociate and the effect of this network fades. In turn, the
response of the entropic chains becomes dominant. The unloading behavior
is governed by the elastic network and the recovery is the result
of the reformation of the intermolecular bonds. 

Within the context of silk modeling, this approach covers an intermediate
scale between those studies that describe the material from a mainly
macroscopic perspective where the microscopic deformation mechanisms
are embedded in a set of constants \citep{Porter2005}, and those
focused on the atomistic and/or molecular level \citep{Brough2024}.
These latter analyses offer essential information on silk-based system
built from the supramolecular organization of fibroin proteins \citep{Venkatesan2019,Du2023},
but do not tend to be readily applicable to the macroscopic scale
of the material.

Thus, the main merit of the present approach stems from its ability
to explicitly capture the evolution of the network configuration during
cyclic loading, including the accumulation of plastic strain and the
reformation of bonds during relaxation periods, while allowing to
establish a connection with the microstructural details of the material.
To demonstrate the merit of the proposed model, we compare its predictions
to two sets of experimental data from \emph{Argiope bruennichi} dragline
silk with an intermediate value of $\alpha^{\star}=0.38$ \citep{Blamires2023}:
(1) continuous uniaxial loading to failure and (2) uniaxial cyclic
loading. We show that the model agrees with the experimental findings.
Needless to say, it is necessary to validate the predictions of the
model with a much wider set of spider species. However, the selection
of this intermediate value of the $\alpha^{*}$ within the context
of the Spider Silk Standardization Initiative (S3I) \citep{Garrote2020}
suggests that the obtained results can be extrapolated, at the very
least, to those spider silk fibers with values of $\alpha^{*}$ not
very far from $\alpha^{*}=0.38$. Currently, there are 14 species
with $\alpha^{*}$ values between 0.3 and 0.5 in S3I.

The paper is organized as follows: we begin by delineating the mechanisms
that govern the elasto-plastic response of the fiber in Section \ref{sec:Deformation-mechanisms}.
Next, the unloading and the relaxation mechanisms are discussed and
modeled in Section \ref{sec:Unloading-and-recovery}. Section \ref{sec:Comparison-to-experiments}
compares the model predictions to experimental findings. The main
conclusions are summarized in Section \ref{sec:Conclusions}.

\section{Deformation mechanisms \label{sec:Deformation-mechanisms}}

To better understand the underlying mechanisms that govern the response
of spider silk fibers, we develop an energy motivated statistical
mechanics based framework that sheds light on the relations between
microstructural quantities and the overall macroscopic response under
cyclic loading of spider silk fibers. We start with the kinematics
and continue with a description of the elasto-plastic mechanisms that
enables us to predict the fiber response. 

\subsection{Kinematics of the cyclic loading}

\begin{figure}[t]
\begin{centering}
\includegraphics[width=16cm]{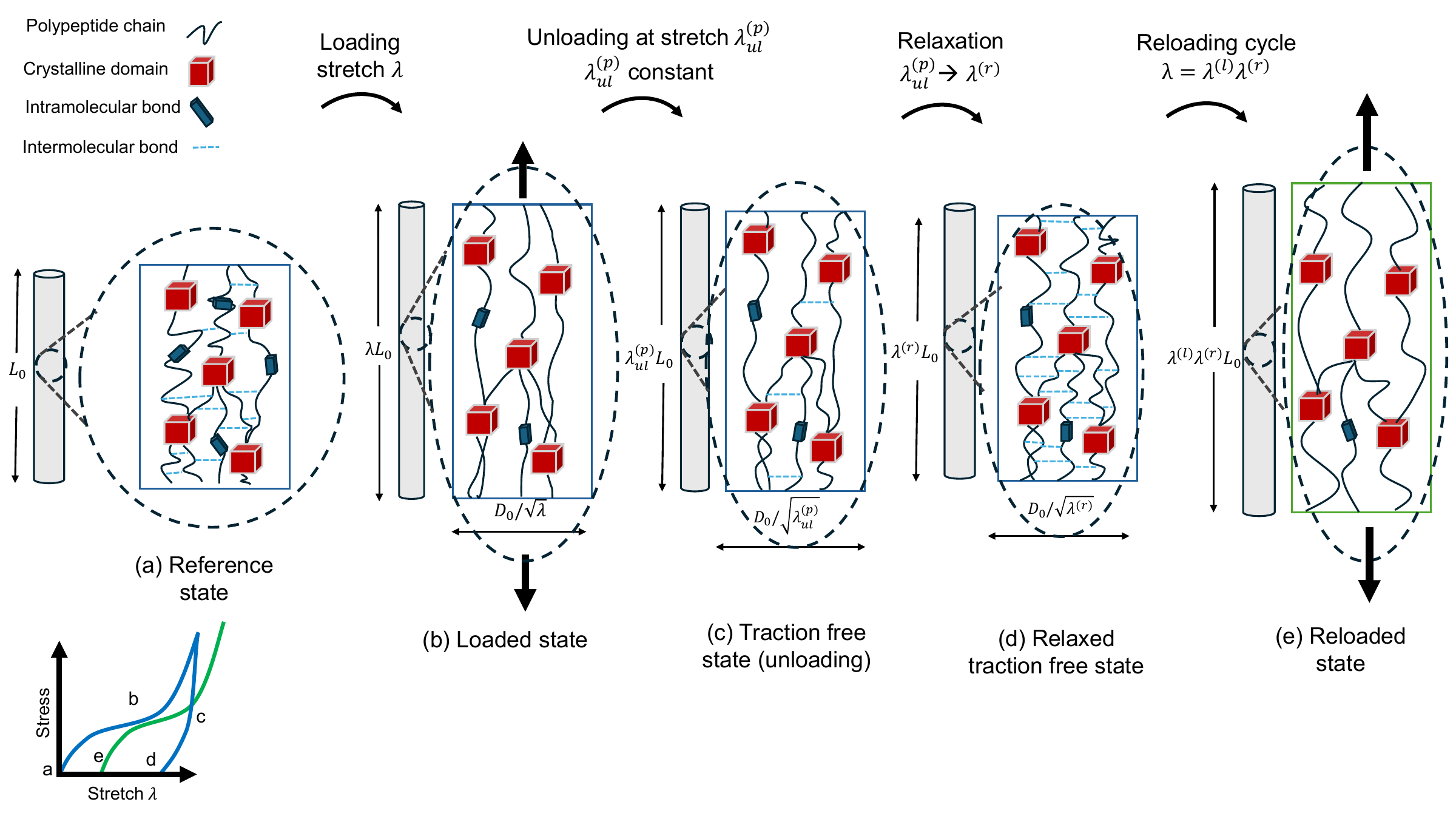}
\par\end{centering}
\caption{Microstructural evolution of the spider silk network during a uniaxial
cyclic loading test: (a) the initial referential fiber, (b) the loaded
state, deformation governed by the dissociation of intermolecular
bonds and entropic chain stretching, (c) the unloaded traction free
state with a residual stretch $\lambda_{ul}^{(p)}$ due to a decrease
in intermolecular bond-density, (d) the relaxed traction free state
in which the intermolecular bonds reform to fix the chains and lock
the fiber in an elongated configuration with a stretch $\ensuremath{\lambda^{(r)}}$,
and (e) the reloading of the fiber, which is characterized by a stiffening
due to the reformation of the bonds and the higher chain stretches.
\label{fig:cyclic_loading}}
\end{figure}

The mechanical response of spider silk fibers under cyclic loading
is characterized by hysteresis and residual strain, with plasticity
mechanisms that lead to irreversible microstructural changes due to
the applied uniaxial loading \citep{veho&etal07BJ,Jiang2023}. To
understand the underlying mechanisms that induce and govern plasticity,
one must first carefully examine the microstructure of the silk fibers.

In the reference undeformed state, shown in Fig. \ref{fig:cyclic_loading}a,
the fibers comprise a network of polypeptide chains that are connected
by crystalline domains (stacked $\beta$-sheets), which serve as permanent
cross-linkers \citep{du&etal11AFM}. The chains interact with each
other through a series of intermolecular hydrogen bonds that restrict
mobility. Furthermore, the chains comprise intramolecular bonds that,
when stretched, reveal a hidden length that extends the contour length
of the chains \citep{du&etal06BJ,Oroudjev2002,du&etal11AFM,Olive2024,Olive2025}.
The fiber is subjected to uniaxial cyclic loading, i.e. it is loaded
from an initial reference configuration, unloaded to a traction free
state, which is characterized by a residual stretch, and then loaded
again. 

In the following, we characterize the kinematics of the uniaxial cyclic
loading. To this end, we define five states, schematically shown in
Fig. \ref{fig:cyclic_loading}: (a) the reference state, (b) the loaded
state, (c) the unloaded and traction free state, (d) the relaxed traction
free state, and (e) the reloaded state. Fig. \ref{fig:cyclic_loading}
also illustrates the microstructural evolution of the fiber during
the cyclic loading process. 

The reference state, depicted in Fig. \ref{fig:cyclic_loading}a,
characterizes the reference state of a glassy silk fiber comprising
$\chains$ chains per unit referential volume. The referential length
and diameter of the fiber are $\reflen$ and $\refdiam$, respectively. 

Next, the fiber is subjected to an external tensile force that leads
to uniaxial extension. The length in the uniaxially loaded state is
$\stretch\reflen$ (see Fig. \ref{fig:cyclic_loading}b). Following
experimental observations \citep{grub&jeli97Macromolecules,guin&etal06Biomacromolecules},
we assume that the fiber is incompressible and accordingly the deformed
diameter is $\refdiam/\sqrt{\stretch}$. We point out that the loading
process induces plasticity, which is described in the following. The
initial deformation of the fiber is enabled by two mechanisms: (1)
the distortion of the intermolecular bonds and (2) the entropic stretching
of polypeptide chains. The former provides the initial stiffness of
the network and is the main source of plasticity - as the loading
increases, the bonds distort and gradually dissociate up to the yield
stress. The dissociation of the bonds transfers the local forces to
the chains \citep{Olive2024,tomm&etal10BJ,elic&etal11JMBBM,du&etal06BJ}.
As the loading increases, bonds may reform and re-break at higher
stretch state, with an overall decrease in the bond-density. Accordingly,
the plasticity accumulates. The entropic stretching of the chains
is associated with the elastic deformation of the fiber. 

To end the loading cycle, the fiber is unloaded to a traction free
state. In this configuration, the length of the fiber is $\stretchpul\reflen$
and, due to the incompressibility, its diameter is $\refdiam/\sqrt{\stretchpul}$,
where $\stretchpul$ is the residual stretch (see Fig. \ref{fig:cyclic_loading}c).
The unloading of the fibers is governed primarily by the entropic
shortening of the elastic polypeptide chains. Due to the decrease
in bond-density during loading, the chains are subjected to less constraints
(or, alternatively, have a higher degree of mobility). As a result,
the network does not recover its referential dimensions upon unloading
and a new traction free configuration that is characterized by a plastic
residual stretch is reached. 

This marks the onset of the relaxation period, in which the chains
reorganize and the intermolecular bonds reform to effectively fix
the conformation of the chains \citep{Jiang2023a,Hong2025}. We denote
the length and the diameter of the fiber in the relaxed traction free
state by $\stretchr\reflen$ and $\refdiam/\sqrt{\stretchr}$, respectively,
as shown in Fig. \ref{fig:cyclic_loading}d. The work of \citet{Jiang2023}
showed that in the first 3 cycles there is minor relaxation ($\stretchr/\stretchpul\sim0.96$),
which becomes completely negligible at higher cycles. It is also pointed
out that the fibers may exhibit a slightly higher degree of alignment,
and therefore a larger number of intermolecular bonds form in order
to counteract the entropic mechanisms that aim to shorten the chains
\citep{Olive2024}. 

Subsequent loadings are typically characterized by a higher stiffness
and yield stress \citep{Jiang2023,veho&etal07BJ,Patil2022}. This
corresponds to the relaxation phase of the fiber - the reformation
of the intermolecular bonds fixes the chains at a higher stretch state
and potentially higher alignment. In the reloaded state, we write
the stretch with respect to the initial reference configuration as
$\stretch=\stretchl\stretchr$, where $\stretchl$ is the ratio between
the length of the relaxed and the deformed fiber. Accordingly, one
can write the deformed length and diameter of the fiber after reloading
$\stretchl\stretchr\reflen$ and $\refdiam/\sqrt{\stretchl\stretchr}$,
respectively (Fig. \ref{fig:cyclic_loading}e). 

Before proceeding, it is important to note that the state of the fiber
at any given configuration can be described by the residual stretch
$\stretchpul=\stretchpul\left(\stretchul\right)$ and the relaxation
stretch $\stretchr=\stretchr\left(\stretchul\right)$, where $\stretch_{ul}$
corresponds to the maximum stretch reached in the previous cycle.
In the reference (initial) configuration, one can set $\stretchpul=\stretchr=1$. 

\subsection{Macroscopic deformation}

To characterize the elasto-plastic behavior during loading, we define
a coordinate system $\left\{ \xh,\yh,\zh\right\} $, with $\xh$ denoting
the fiber direction. Accordingly, the deformation gradient from a
relaxed traction free state to a loaded configuration can be written
as
\begin{equation}
\defgradT=\stretchl\xh\otimes\xh+\frac{1}{\sqrt{\stretchl}}\left(\yh\otimes\yh+\zh\otimes\zh\right).
\end{equation}
It is emphasized that the stretch with respect to the initial configuration
(i.e. the ratio between the length of the fiber in a deformed and
the initial reference configuration) is $\stretch=\stretchr\stretchl$. 

\begin{figure}[t]
\begin{centering}
\includegraphics[width=16cm]{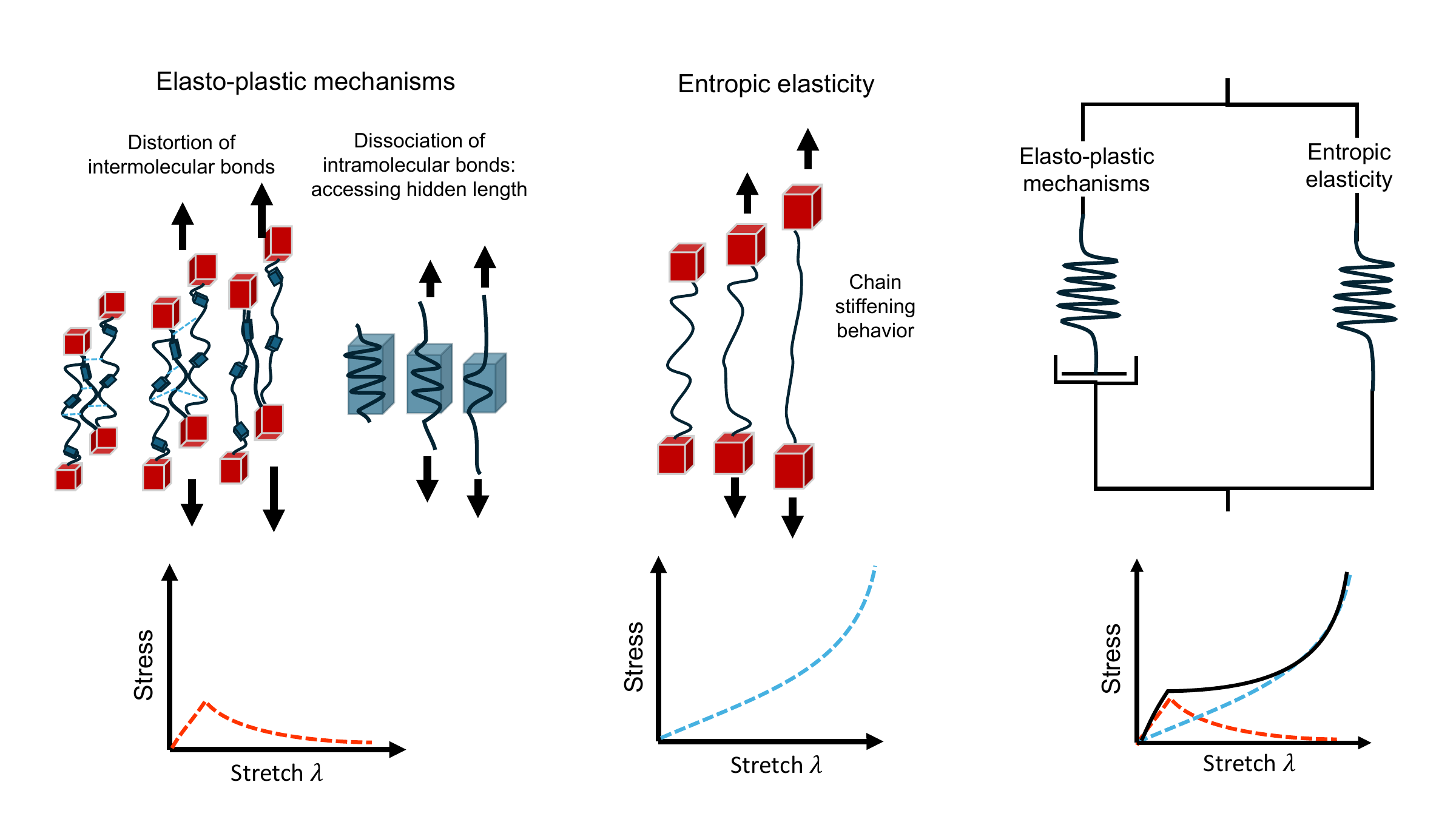}
\par\end{centering}
\caption{Schematic decomposition of the macroscopic mechanical response of
spider silk fibers into two networks connected in parallel: (left)
a network that captures the elasto-plastic mechanisms, governed by
the dissociation of intermolecular and intramolecular bonds, and (center)
a network of entropic chains that stretch elastically. (Right) The
two springs connected in parallel. The bottom of the figure plots
the stress-stretch curves corresponding to the elasto-plastic network
(red), the entropic network (blue), and the macroscopic stress-strain
curve (black), which is the summation of the two contributions. \label{fig:parallel_springs}}
\end{figure}

In the following, we propose a model for the elasto-plastic mechanisms
that govern the loading cycle. Recall that the overall macroscopic
response of the spider silk fiber depends on the intermolecular bonds
and the entropic elasticity of the polypeptide chains. We assume that
the plasticity is dominated by the distortion of the intermolecular
bonds and their dissociation while the response of the chains is entropic
and elastic. Following previous works \citep{Bergstroem1998,flor42JCP,Bernard2016,Cho2015,Fontenele2023,Anand2009},
we decouple the macroscopic response into two networks that are connected
in parallel, as shown in Fig. \ref{fig:parallel_springs}. Here, the
left side represents the network of intermolecular bonds that distort,
thereby allowing the chains to slide past each other to yield deformation.
The bonds can be considered as frictional elements that yield and
dissipate energy, corresponding to the initial linear response. In
addition, we plot the intramolecular bonds that can extend to increase
the chain length, as discussed in previous works \citep{du&etal11AFM,Olive2024}.
In parallel, the center of Fig. \ref{fig:parallel_springs} shows
a network of entropic chains that deform elastically. The right-hand
side of Fig. \ref{fig:parallel_springs} plots the two springs connected
in parallel, with the stress-stretch curves corresponding to the elasto-plastic
network (red), the entropic network (blue), and the macroscopic stress-strain
curve (black), which is the summation of the two contributions.

Accordingly, we can write the deformation gradient $\defgradT=\defgradTeb\defgradTpb=\defgradTen$,
where the superscripts $\left(b\right)$ and $\left(n\right)$ denote
the elasto-plastic network of intermolecular bonds and the elastic
network of chains, respectively. It is convenient to recast the latter
in terms of the axial stretch, 
\begin{equation}
\stretchl=\stretcheb\stretchpb=\stretchen,\label{eq:stretch_l_definition}
\end{equation}
where $\stretcheb$ and $\stretchen$ denote the elastic stretch components
in the bond and the chain networks, respectively, and $\stretchpb$
is the plastic stretch. 

The total energy-density can be written as the sum of the energy-densities
due to the distortion of the intermolecular bonds $\energyb\left(\stretcheb\right)$
and the distortion of the network $\energyn\left(\stretchen\right)$.
Specifically, 
\begin{equation}
\energy\left(\stretcheb,\stretchen\right)=\energyb\left(\stretcheb\right)+\energyn\left(\stretchen\right),
\end{equation}
where $\stretcheb=\stretch/\stretchpb$ (see Eq. \ref{eq:stretch_l_definition}).
Accordingly, the true stress is
\begin{equation}
\stress\left(\stretch\right)=\stressb\left(\stretcheb\right)+\stressn\left(\stretchen\right),\label{eq:stress}
\end{equation}
where $\stressb=\left(\partial\energyb/\partial\stretcheb\right)\stretcheb$
and $\stressn=\left(\partial\energyn/\partial\stretchen\right)\stretchen$
are the stress due to the intermolecular bonds and  the network of
chains, respectively. 

\subsection{The elasto-plastic response}

To account for the stress due to the network of bonds, we assume the
energy-density 
\begin{equation}
\energyb\left(\stretcheb\right)=\frac{\E}{2}\ln\left(\stretcheb\right)^{2},
\end{equation}
and therefore the stress is 
\begin{equation}
\stressb\left(\stretcheb\right)=\E\ln\left(\stretcheb\right).\label{eq:stress_elasto-plastic}
\end{equation}

To account for the plasticity of the network, which stems from the
dissociation of the intermolecular bonds, we define the yield surface
\begin{equation}
\yieldsurface=\stressb-\yieldstress\left(\stretchpb\right),
\end{equation}
where $\yieldstress\left(\stretchpb\right)$ is the yield stress,
given as a function of the accumulating plastic strain. As the loading
increases beyond the yield stress $\yieldstressf>0$, the intermolecular
bonds gradually dissociate and the yield stress decreases. We point
out that the yield stress depends on the plastic strain, and it is
convenient to express $\yieldstressf=\yieldstressf\left(\stretchr\right)$
in terms of the plastic stretch in the relaxed configuration.

Once most of the bonds break, the deformation of the fiber is governed
by entropic elasticity. To capture this effect, we define the yield
stress
\begin{equation}
\yieldstress\left(\stretchpb\right)=\yieldstressf\left(\stretchpb\right)\exp\left(-\yieldc\,\plastic\right),
\end{equation}
where $\yieldc\ge0$ is the exponential decay parameter that accounts
for the ``rate'' of bond dissociation and $\plastic\ge0$ is the
plastic multiplier that accounts for the accumulated plastic strain.
Note that as the physical intermolecular bonds dissociate, the stress
$\stressb\rightarrow0$. 

To determine the evolution of the plastic stretch, we consider the
flow rule 
\begin{equation}
\dot{\stretchpb}=\dot{\plastic}\,\stretchpb,
\end{equation}
where $\dot{\plastic}\ge0$ is the plastic multiplier rate. 

\subsection{Elastic chain network }

\global\long\def\chainTref{\mathbf{R}}%
\global\long\def\chainlenref{R}%
\global\long\def\dirTref{\hat{\mathbf{R}}}%

\global\long\def\chainTcur{\mathbf{r}}%
\global\long\def\chainlencur{r}%
\global\long\def\dirTcur{\hat{\mathbf{r}}}%

\global\long\def\links{n}%
\global\long\def\contourlen{L_{c}}%
\global\long\def\rat{\rho}%
\global\long\def\ratinit{\alpha}%

\global\long\def\lang{\tau}%
\global\long\def\stressTc{\boldsymbol{\sigma}_{c}}%
\global\long\def\stressTn{\boldsymbol{\sigma}^{\left(n\right)}}%
\global\long\def\pressure{p}%

To determine the stress associated with the network of polypeptide
chains, we model the chains as freely jointed chains. To this end,
consider a network with $\chains$ chains per unit referential volume,
where each chain comprises $\links$ repeat units and has a contour
length $\contourlen$. In the undeformed configuration, the end-to-end
vector of the $i$-th chain is $\chainTref^{\left(i\right)}=\chainlenref\dirTref^{\left(i\right)}$,
where $\chainlenref=\ratinit\contourlen$ and $\dirTref^{\left(i\right)}$
are the end-to-end distance and direction, respectively. Here, $\ratinit$
denotes the ratio between the end-to-end distance in the relaxed state
and the contour length. Following common practice, we assume that
the chain experiences the macroscopic deformation gradient $\defgradTen$
such that the deformed end-to-end vector of the $i$-th chain is $\chainTcur^{\left(i\right)}=\defgradTen\chainTref^{\left(i\right)}=\chainlencur^{\left(i\right)}\dirTcur^{\left(i\right)}$,
with $\chainlencur^{\left(i\right)}$ and $\dirTcur^{\left(i\right)}$
as the deformed end-to-end distance and direction, respectively. The
stress associated with the chain can be written as \citep{cohe&eise19AB,cohe&etal21BMM,Olive2024}
\begin{equation}
\stressTc^{\left(i\right)}=\boltzmann T\links\,\rat^{\left(i\right)}\,\lang\left(\rat^{\left(i\right)}\right)\,\dirTcur^{\left(i\right)}\otimes\dirTcur^{\left(i\right)},
\end{equation}
where $\boltzmann$ is the Boltzmann constant, $T$ is the temperature,
$\rat=\chainlencur/\contourlen$ is the ratio between the deformed
end-to-end distance and the contour length of a chain, and $\lang=\lang\left(\rat\right)$
is determined from the Langevin function $\rat=\coth\lang-1/\lang$.
It is useful to employ the approximation $\lang\approx\rat\left(3-\rat^{2}\right)/\left(1-\rat^{2}\right)$
\citep{cohe91RA}.

The macroscopic stress of the chain network is given by
\begin{equation}
\stressTn=\sum_{All\,chains}\stressTc^{\left(i\right)}-\pressure\mathbf{I}=\chains\left\langle \stressTc\right\rangle -\pressure\mathbf{I},\label{eq:total_stress_chain_network}
\end{equation}
where $\left\langle \stressTc\right\rangle $ is the average stress
on a chain and $\pressure$ is a pressure-like term that ensures the
incompressibility of the network. Once the overall stress is determined,
we employ the boundary conditions $\stressTn\yh\cdot\yh=\stressTn\zh\cdot\zh=0$
to determine the pressure term $\pressure=\chains\left\langle \stressTc\right\rangle \yh\cdot\yh=\chains\left\langle \stressTc\right\rangle \zh\cdot\zh$
and, consequently, the uniaxial stress 
\begin{equation}
\stressn=\chains\left\langle \stressTc\right\rangle \xh\cdot\xh-\pressure.\label{eq:uniaxial_stress_chain_network}
\end{equation}

\section{Unloading and recovery mechanisms \label{sec:Unloading-and-recovery}}

\begin{figure}[t]
\begin{centering}
\includegraphics[width=14cm]{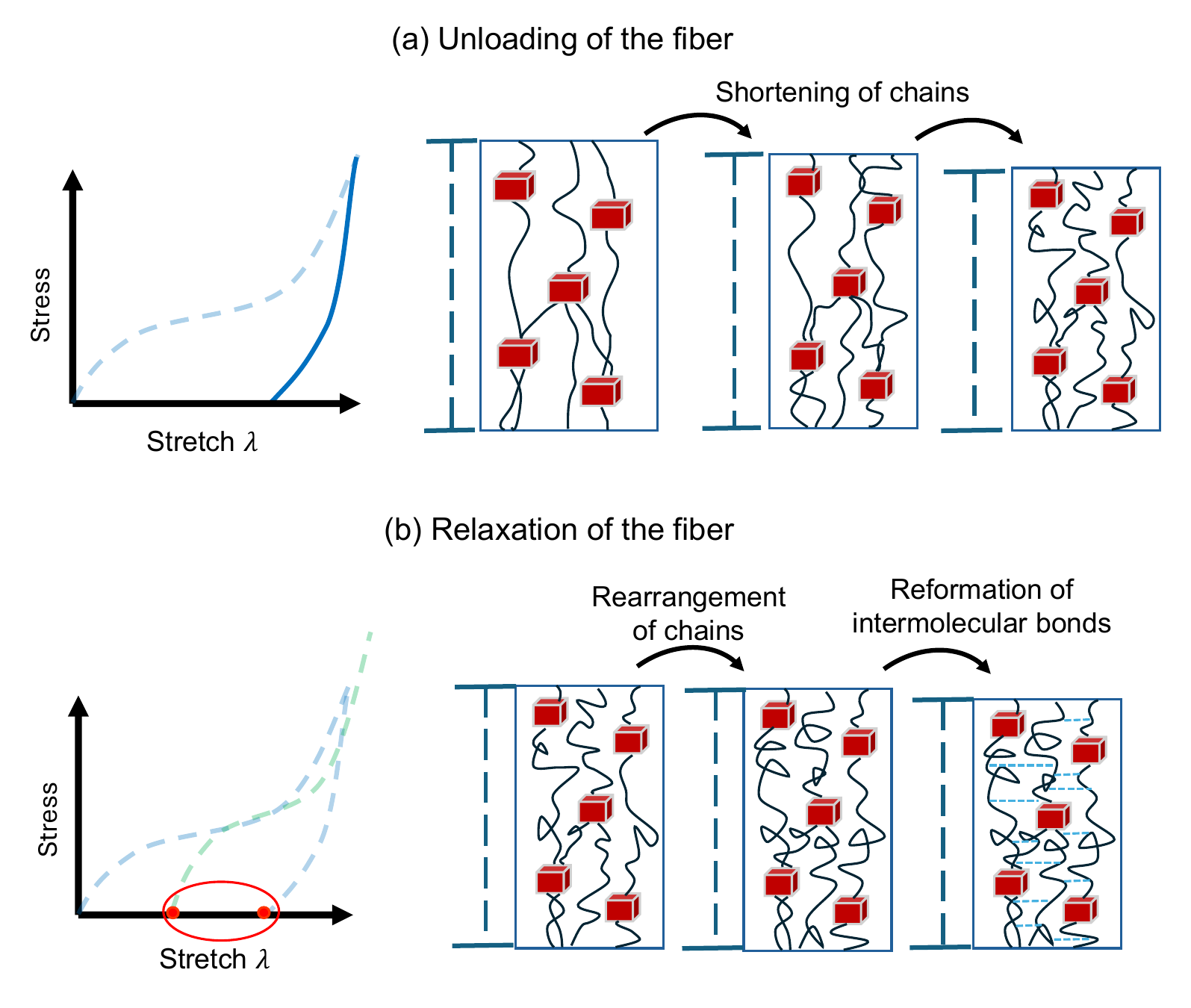}
\par\end{centering}
\caption{(a) The unloading and (b) the relaxation mechanisms governing the
recovery of the spider silk fiber. The unloading phase is characterized
by entropic chain shortening and coiling, leading to a residual stretch.
The relaxation phase in the traction free state involves the reorganization
of the chains in the network and the reformation of intermolecular
bonds, which fix the microstructure and lead to a stiffening in the
subsequent loading cycle. \label{fig:Unloading_evolution}}
\end{figure}

Experiments show that the behavior of the silk fiber changes from
cycle to cycle during a cyclic loading \citep{Jiang2023}. To understand
this phenomenon, we carefully examine the unloading and the relaxation
mechanisms that govern the recovery of the fiber. First, recall that
once the yield point is reached, the intermolecular bonds dissociate
and their density decreases. Consequently, the response during loading
is governed by the entropic elasticity of the polypeptide chains.
As a result of the decrease in the number of intermolecular bonds,
the unloading is also enabled by the entropic shortening of the chains.
Furthermore, a traction free state is reached at stretches that are
larger than those in the initial reference configuration due to the
irreversible loss of the bond, leading to a growing residual stretch
$\stretchpul$. This process is illustrated in Fig. \ref{fig:Unloading_evolution}a. 

Next, the fiber is allowed to relax and recover in a traction free
state. During this recovery process, the fiber is allowed to ``heal''
through the reorganization of the chains and the reformation of the
intermolecular bonds, which fix the microstructure in a ``new''
configuration. This is schematically shown in Fig. \ref{fig:Unloading_evolution}b.
In the experiments of \citet{Jiang2023}, the fiber was held at zero
stress for 20 minutes and an increase in the stiffness and the yield
stress was observed. These two effects stem from (1) a higher stretching
of polypeptide chains (captured by the parameter $\ratinit$ in our
model) and (2) an increase in intermolecular bond-density, which is
required to counteract the entropically motivated shortening of the
chains \citep{Olive2024}. We point out that a higher degree of chain
alignment is also plausible. This behavior persists as additional
cycles are performed. 

The recovery process highlights the adaptive nature of the spider
silk network. The relaxation period allows the fiber to minimize its
energy not by macroscopic contraction, but by microscopic reorganization,
effectively ``locking'' the residual stretch into a new stable equilibrium.
Furthermore, since the reformation of hydrogen bonds occurs within
a pre-aligned chain network, the microstructure is fixed in a preferred
molecular extension and orientation. As discussed by \citet{Olive2024},
the additional extension of the chains leads to a higher local energy
state and therefore a higher density of intermolecular bonds is required
to counteract the entropic forces working towards contracting the
chains. This directly results in a stiffer fiber with a higher yield
stress. 

\section{Comparison to experiments \label{sec:Comparison-to-experiments}}

\begin{table}
\centering
\caption{A summary of the model parameters and their significance. \label{tab:summary_of_parameters}}

\centering{}%
\begin{tabular}{cc}
\toprule 
Parameter & Physical meaning\tabularnewline
\midrule
\midrule 
$\stretch,\stretchul$ & The stretch and the stretch at which the fiber is unloaded\tabularnewline
\midrule 
$\stretchpul,\stretchr$ & Plastic stretch, relaxation stretch\tabularnewline
\midrule 
$\stretchl$ & Stretch with respect to relaxed state\tabularnewline
\midrule 
$\stretcheb,\stretchen$ & Elastic stretch of bond and chain networks\tabularnewline
\midrule 
$\stretchpb$ & Plastic stretch of bond network\tabularnewline
\midrule 
$\stressb,\stressn$ & Stress contributions from bond and chain networks\tabularnewline
\midrule 
$\yieldsurface,\yieldstress$ & Yield surface, yield stress\tabularnewline
\midrule 
$\yieldc,\plastic$ & Exponential decay parameter, plastic multiplier\tabularnewline
\midrule 
$N_{0}$ & Chain-density \tabularnewline
\midrule 
$\links,\contourlen$ & Number of repeat units and contour length of chain\tabularnewline
\midrule 
$\ratinit$ & The ratio between the end-to-end distance in the relaxed state and
the contour length\tabularnewline
\bottomrule
\end{tabular}
\end{table}

To validate the model, we compare its predictions to the experimental
findings of \citet{Jiang2023}, who performed cyclic loading tests
on a supercontracted major ampullate fiber from \emph{Argiope bruennichi}.
It is emphasized that the experiments were performed on supercontracted
fibers - spider silk fibers were immersed in water for 30 minutes
and left to dry overnight. During this process, the alignment of the
polypeptide chains was lost and the resulting dehydrated fiber comprises
a glassy network of (roughly) randomly oriented and uniformly distributed
chains \citep{plaz&etal06JPS,fazi&etal22JMPS,cohe&etal21BMM,Olive2024,blac&etal09JEB}.
Therefore, to integrate from the chain to the macroscopic level in
the calculation of the stress (Eq. \ref{eq:total_stress_chain_network}),
we employ the well-known micro-sphere technique, which is summarized
in Appendix \ref{sec:micro-sphere}. This method was successfully
used to capture the response of biological networks \citep{cohe&eise19AB,cohe&mcme19JMPS,Olive2024}.
We point out that alignment can be captured by considering an orientation
distribution function and additional chain directions, as shown in
the work of \citet{Olive2024}.

For convenience, the main model parameters are listed in Table \ref{tab:summary_of_parameters}.

\subsection{Stiffness, yield stress, and relaxation stretch}

\global\long\def\Eref{E_{0}}%
\global\long\def\Eslope{\zeta_{E}}%
\global\long\def\yieldref{\yieldstress^{\left(0\right)}}%
\global\long\def\yieldinf{\yieldstress^{\left(\infty\right)}}%
\global\long\def\yieldslope{\zeta_{y}}%
\global\long\def\stretchrslope{\zeta_{r}}%
\global\long\def\aref{\ratinit^{\left(0\right)}}%
\global\long\def\ainf{\ratinit^{\left(\infty\right)}}%
\global\long\def\aslope{\zeta_{\alpha}}%

We begin by examining the dependence of the Young's modulus and the
yield stress on the loading history of the fiber. \citet{Jiang2023}
cyclically stretched three different fibers from \emph{Argiope bruennichi}
and measured the Young's modulus and the yield stress as a function
of the relaxation stretch $\stretchr$ (see Figs. 6 and 7 in that
paper). 

\begin{figure}[t]
\hspace*{\fill}(a) \includegraphics[width=6.5cm]{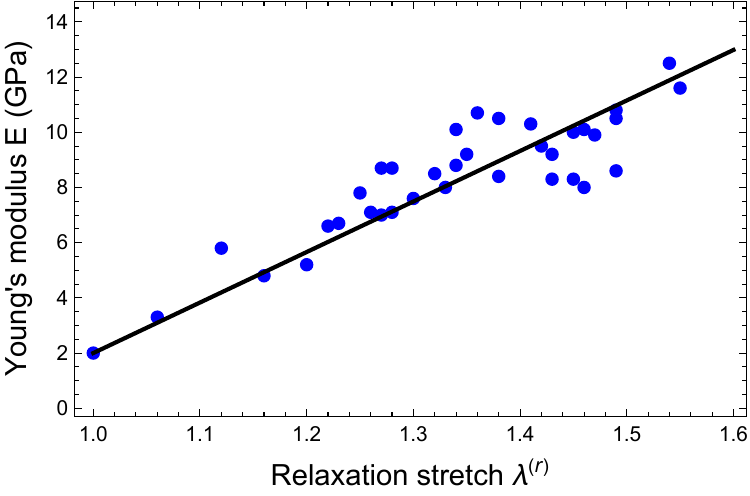}~~~(b)
\includegraphics[width=6.5cm]{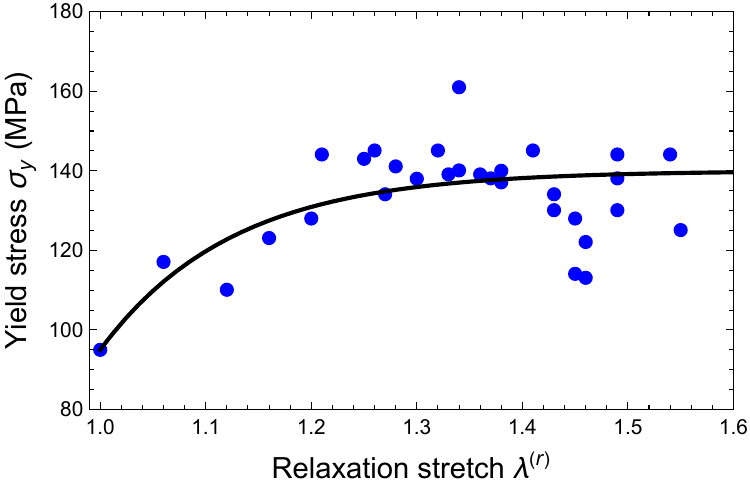}\hspace*{\fill}

\hspace*{\fill}(c) \includegraphics[width=6.5cm]{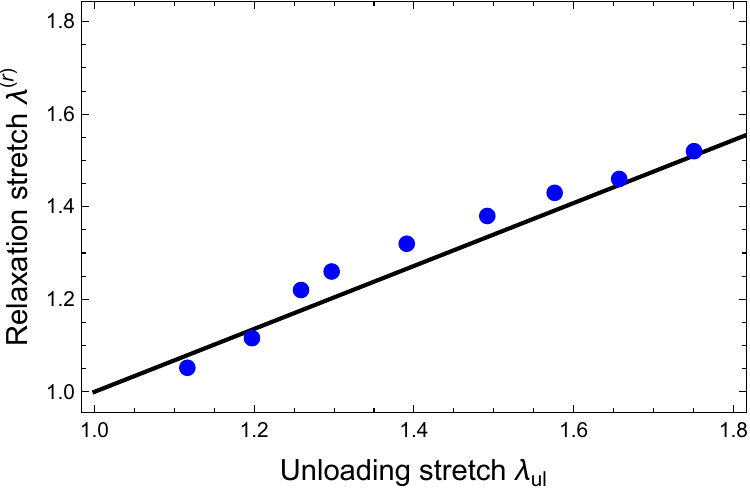}\hspace*{\fill}

\caption{(a) The Young's modulus $E$ (Eq. \ref{eq:E_eq}) and (b) the yield
stress $\protect\yieldstress$ (Eq. \ref{eq:yield_eq}) as a function
of the relaxation stretch $\protect\stretchr$. (c) The relaxation
stretch $\protect\stretchr$ (Eq. \ref{eq:stretch_r_eq}) as a function
of the unloading stretch $\protect\stretchul$. The continuous curves
correspond to proposed expressions and the circle marks denote the
experimental findings of \citet{Jiang2023}. \label{fig:fiber_properties}}
\end{figure}

With the aim of understanding the overall effect and the resulting
trend from cyclic loading, we propose expressions to fit the experimental
measurements of \citet{Jiang2023}. The Young's modulus exhibits a
linear trend such that 
\begin{equation}
\E=\Eref+\Eslope\left(\stretchr-1\right),\label{eq:E_eq}
\end{equation}
where $\Eref$ is the Young's modulus of the virgin fiber and $\Eslope$
is the slope. The yield stress initially increases with the cycles
and then attenuates to reach a limiting value. To capture this, we
propose the expression
\begin{equation}
\yieldstress=\yieldinf-\left(\yieldinf-\yieldref\right)\exp\left(-\yieldslope\left(\stretchr-1\right)\right),\label{eq:yield_eq}
\end{equation}
where $\yieldref$ and $\yieldinf$ denote the initial and the limiting
yield stress, respectively, and $\yieldslope$ is a parameter that
captures the dependence of $\yieldstress$ on the relaxation stretch
$\stretchr$. 

Figs. \ref{fig:fiber_properties}a and \ref{fig:fiber_properties}b
plot the Young's modulus $E$ (Eq. \ref{eq:E_eq}) and the yield stress
$\yieldstress$ (Eq. \ref{eq:yield_eq}) as a function of the relaxation
stretch $\stretchr$, respectively. The circle marks denote the experimental
data and the fitted parameters for the trend lines are $\Eref=2\,\mathrm{GPa}$,
$\Eslope=18.3\,\mathrm{GPa}$, $\yieldref=95\,\mathrm{MPa}$, $\yieldinf=140\,\mathrm{MPa}$,
and $\yieldslope=8$. It is shown that while the results exhibit a
clear trend, the variance and the variability between fibers is inherent.
In fact, such phenomenon is common in biological materials, and previous
works highlighted the variability in mechanical properties between
spider species and between fibers from the same spider species \citep{Madsen1999,PerezRigueiro2003,voll00RMB,guin&etal05JEB,elic&etal11JMBBM}. 

Another important measurable quantity of interest is the dependence
of the relaxation stretch $\stretchr$ on the unloading stretch $\stretchul$
from the previous cycle. This relation can be approximated via
\begin{equation}
\stretchr=1+\stretchrslope\left(\stretchul-1\right),\label{eq:stretch_r_eq}
\end{equation}
where $\stretchrslope$ is the slope. Fig. \ref{fig:fiber_properties}c
depicts this relation, where $\stretchrslope=0.68$ is used, for the
nine cycles shown in Fig. 4 of \citet{Jiang2023}. 

\subsection{The response to cyclic loading}

To capture the overall behavior of the fiber under cyclic loading,
one must define the ratio $\ratinit$ between the end-to-end distance
in the relaxed state and the contour length. While it is difficult
to measure this quantity, we expect the length of the chain to achieve
a limiting value as additional stretching cycles are exerted. Accordingly,
we propose the relation $\ratinit=\ainf-\left(\ainf-\aref\right)\exp\left(-\aslope\left(\stretchr-1\right)\right)$,
where $\aref=0.47$ and $\ainf=0.78$ are the referential and the
limiting ratios and $\aslope=7$ is an exponential prefactor that
describes the dependence of $\ratinit$ on $\stretchr$. In addition,
we set $\links=6$ to account for the limiting extensibility of the
fiber, the chain density $\chains$ such that $\chains\boltzmann T=0.1\,\mathrm{GPa}$,
and $\yieldc=5$. 

\begin{figure}[t]
\hspace*{\fill}\includegraphics[width=6.5cm]{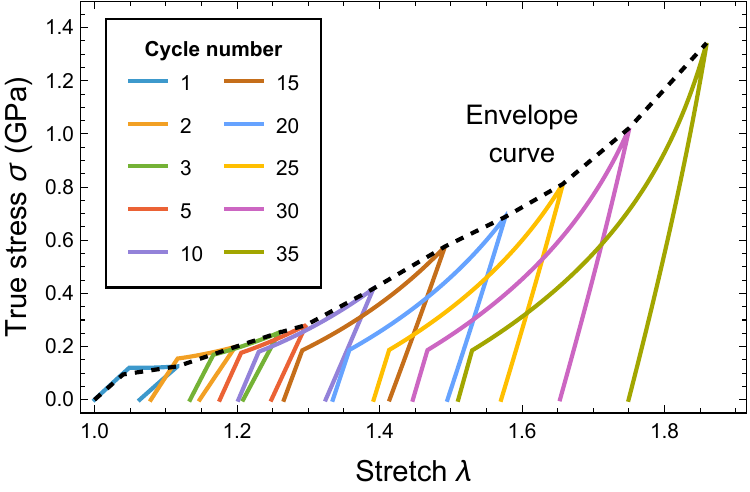}\hspace*{\fill}

\caption{The true (Cauchy) stress $\protect\stress$ as a function of the stretch
$\protect\stretch$ for a fiber subjected to cyclic loading. \label{fig:cyclic_loading_full}}
\end{figure}

Fig. \ref{fig:cyclic_loading_full} plots the true (Cauchy) stress
$\stress$ as a function of the stretch $\stretch$ for a fiber subjected
to cyclic loading. The dashed line denotes the envelope curve, which
is obtained by interpolating the values between the maximum values
of the true stress reached in each cycle. This plot qualitatively
and quantitatively agrees with the experiments shown in Fig. 4 of
\citet{Jiang2023}. The differences between the model predictions
and the data stem from the variance-induced errors of Eqs. \ref{eq:E_eq},
\ref{eq:yield_eq}, and \ref{eq:stretch_r_eq}. 

To demonstrate the ability of the model to predict the response for
a given cycle, we independently fit our model to three types of experiments
from \citet{Jiang2023}: (1) continuous loading, (2) the loading cycles
2 and 3, and (3) the loading of cycles 20 and 30. In all of the simulations
we once again set the number of repeat units in a chain $\links=6$
and the chain density $\chains$ such that $\chains\boltzmann T=0.1\,\mathrm{GPa}$.
The remaining model parameters for the continuous and the cyclic loadings
are fitted to experimental data and listed in Table \ref{tab:model_parameters_cyclic}.
The differences between the fitted values and the proposed expressions
in Eqs. \ref{eq:E_eq} and \ref{eq:yield_eq} are subsequently discussed. 

\begin{table}[t]
\caption{Fitted model parameters for the cyclic loading simulations. \label{tab:model_parameters_cyclic}}

\centering{}%
\begin{tabular}{ccccccc}
\toprule 
 & $\stretchr$ & $\stretchul$ & $\E\,\left(\mathrm{GPa}\right)$ & $\yieldstress\,\left(\mathrm{MPa}\right)$ & $\ratinit$ & $\yieldc$\tabularnewline
\midrule
\midrule 
Loading & $1$ & $-$ & $2$ & $80$ & $0.47$ & $5$\tabularnewline
\midrule 
Cycle 2 & $1.05$ & $1.19$ & $3$ & $84$ & $0.49$ & $5$\tabularnewline
\midrule 
Cycle 3 & $1.12$ & $1.27$ & $4.4$ & $95$ & $0.54$ & $5$\tabularnewline
\midrule 
Cycle 20 & $1.36$ & $1.57$ & $9$ & $130$ & $0.772$ & $15$\tabularnewline
\midrule 
Cycle 30 & $1.46$ & $1.76$ & $10$ & $130$ & $0.778$ & $20$\tabularnewline
\bottomrule
\end{tabular}
\end{table}

\begin{figure}[t]
\hspace*{\fill}(a) \includegraphics[width=6.5cm]{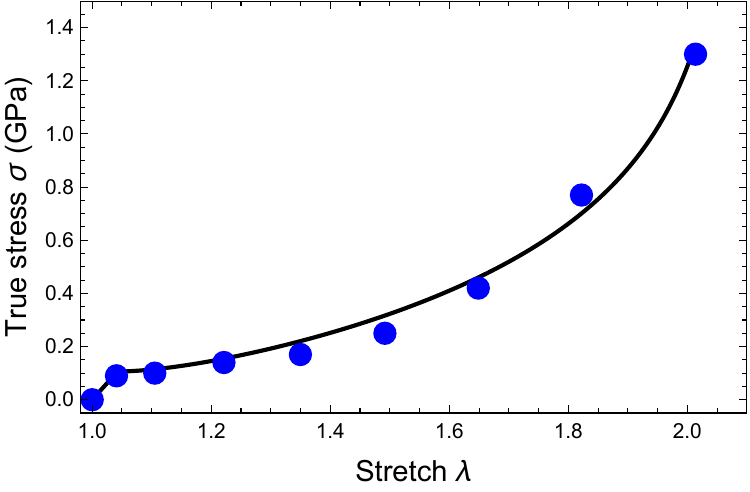}~~~(b)
\includegraphics[width=6.5cm]{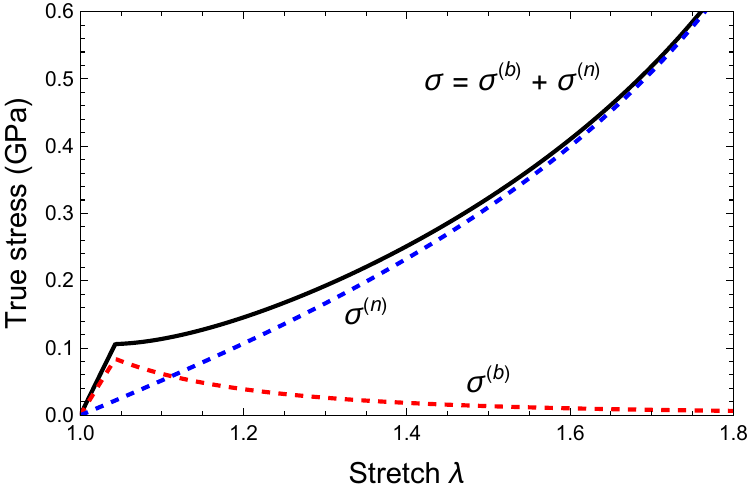}\hspace*{\fill}

\caption{Loading of a spider silk fiber: (a) true stress $\protect\stress$
as a function of the stretch $\protect\stretch$. The continuous curve
corresponds to the model predictions and the circle marks denote the
experimental findings of \citet{Jiang2023}. (b) The stress due to
the distortion of the intermolecular bonds $\protect\stressb$ (Eq.
\ref{eq:stress_elasto-plastic}), the entropic stress $\protect\stressn$
(Eq. \ref{eq:uniaxial_stress_chain_network}), and the total stress
$\protect\stress=\protect\stressb+\protect\stressn$ (Eq. \ref{eq:stress})
as a function of the stretch $\protect\stretch$. \label{fig:results-loading}}
\end{figure}

Fig. \ref{fig:results-loading}a plots the true stress $\stress$
as a function of the stretch $\stretch$ according to the experimental
findings of \citep{Jiang2023} (circle marks) and the proposed model
(continuous black curve). The model agrees with the experimental findings.
To emphasize the contributions of the different mechanisms, Fig. \ref{fig:results-loading}b
plots the stress due to the distortion of the intermolecular bonds
$\stressb$ (Eq. \ref{eq:stress_elasto-plastic}), the entropic stress
due to the deformation of the chains $\stressn$ (Eq. \ref{eq:uniaxial_stress_chain_network}),
and the total stress $\stress=\stressb+\stressn$ (Eq. \ref{eq:stress})
as a function of the stretch $\stretch$. We find that the initial
linear regime is governed by the distortion of the intermolecular
hydrogen bonds (see dashed red curve). Once the yield stress is reached,
most of the bonds dissociate and the load is transferred to the elastic
network of chains (marked by the dashed blue curve). As the external
force increases, the bonds may reform and break with an overall bond-density
that decreases. Beyond the stretch of $\stretch\sim1.35$, the contribution
of the bonds to the overall stress becomes negligible ($<10\%$) and
the behavior is dominated by the entropic elasticity of the polypeptide
chains. 

\begin{figure}[t]
\hspace*{\fill}(a) \includegraphics[width=6.5cm]{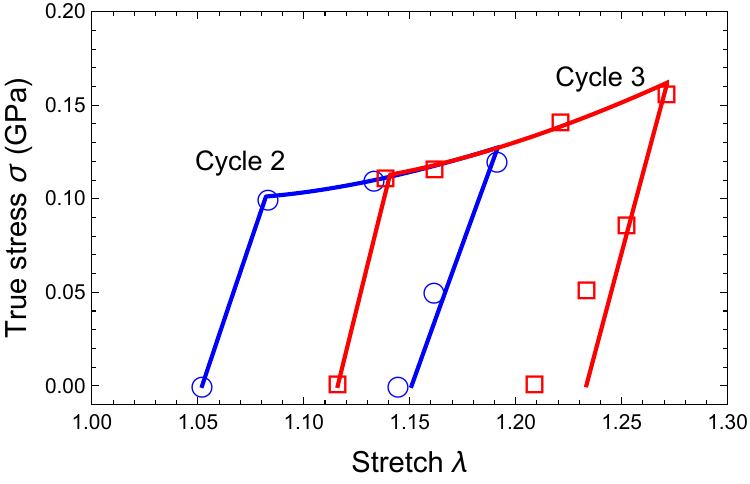}~~~(b)
\includegraphics[width=6.5cm]{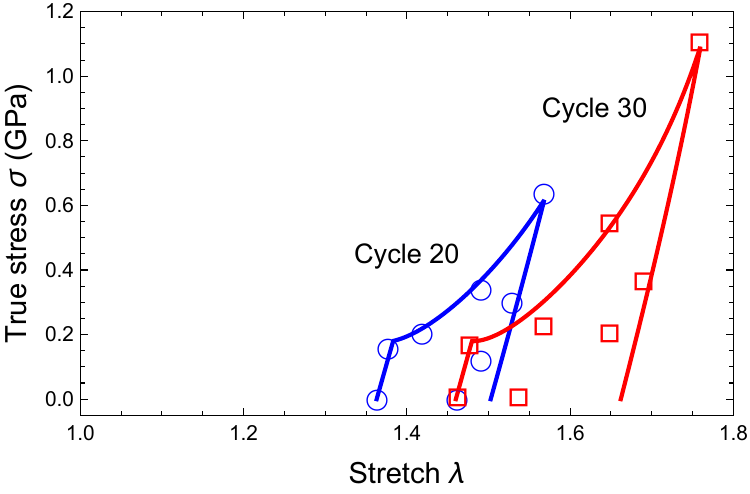}\hspace*{\fill}

\caption{Cyclic loading of a spider silk fiber: true stress $\protect\stress$
as a function of the stretch $\protect\stretch$ for (a) cycles 2
and 3 and (b) cycles 20 and 30 in the work of \citet{Jiang2023}.
The continuous curve corresponds to the model predictions and the
circle marks denote the experimental findings of \citet{Jiang2023}.
\label{fig:Results-cyclic_loading}}
\end{figure}

To demonstrate the robustness of the model and its ability to capture
cyclic loading, we plot the true stress $\stress$ as a function of
the stretch $\stretch$ for cycles 2 and 3 and cycles 20 and 30 from
the experiments of \citet{Jiang2023} in Figs. \ref{fig:Results-cyclic_loading}a
and \ref{fig:Results-cyclic_loading}b, respectively. It is shown
that the model is capable of capturing the response in the different
cycles. The difference between the fitted parameters and Eq. \ref{eq:E_eq}
for the Young's modulus are $<5\%$. However, the proposed Eq. \ref{eq:yield_eq}
for the yield stress fails to capture the fitted values for cycles
2 and 3, with an error of $\sim25\%$. This is the result of the extreme
variability in the experimental findings of these two cycles - while
\citet{Jiang2023} measured a trend of increasing yield stress with
cycles, the decreasing values $\yieldstress\approx117$ and $\yieldstress\approx110$
for cycles 2 and 3, respectively, were reported. The error for cycles
20 and 30 is $<2.5\%$, highlighting the merit of Eq. \ref{eq:yield_eq}.
However, these findings emphasize the need for further experimental
investigations, especially in the context of variability and the first
loading cycles. 

Interestingly, in the early cycles the relaxation leads to a decrease
in stretch from $\stretchpul\sim1.145$ to $\stretchr\sim1.115$.
At higher cycles this effect is attenuated, as shown with cycles 20
and 30. In addition, the reorganization of the chains and the reformation
of the bonds during the relaxation phase lead to a higher stretching
of the chains, as denoted by the change in $\ratinit$. Consequently,
an increase in the Young's modulus $\E$ and the yield stress $\yieldstress$
is observed. 

We also find that the exponential decay parameter $\yieldc$, which
accounts for the rate of dissociation of the intermolecular bonds,
increases with the cycles. To understand this behavior, note that
at higher cycles the chains are more extended in the relaxed traction
free configuration. To counteract the entropic shortening of the chains,
more intermolecular bonds must form. These bonds experience higher
stress due to the entropic force, and are therefore expected to dissociate
at lower external loads. 

\section{Conclusions \label{sec:Conclusions}}

This work presents a model of spider silk fibers developed at an intermediate
scale between the macroscopic approach of mean field theories and
the atomistic/molecular scale of most computer-intensive simulations.
Thus, it establishes insightful connections between the microstructural
features of the material and the observed mechanical behavior while
requiring low computational cost. In this regard, the model provides
a foundational understanding of the underlying mechanisms that govern
the elasto-plastic response of spider silk fibers under cyclic loading.
Specifically, the microstructural origin of the plasticity, the hysteresis,
and the recovery of the fibers was discussed and modeled. We start
by considering the microstructure of the silk, which is made of a
network of polypeptide chains that are connected through crystalline
domains to form a network. Prior to loading, the chains comprise intramolecular
$\beta$-sheets that store ``hidden length'' and are interconnected
by a series of intermolecular bonds that restrict mobility. The application
of an initial external load first distorts the bonds to provide a
linear elastic response. Once the yield stress is reached, the density
of the bonds decreases and the load is transferred to the chains.
Consequently, any additional deformation stems from the entropic stretching
of the polypeptide chains. Therefore, the dissociation of the bonds
is the main cause of plasticity in the fiber. Upon unloading, the
shortening of the polypeptide chains drives the decrease in stretch.
Since there is a significant decrease in the density of the bonds,
a traction free configuration that is associated with a residual plastic
stretch is reached. Subsequently, the fiber relaxes such that the
chains reorganize and the intermolecular bonds reform, thereby ``locking''
the network configuration in place. This process yields an increase
in stiffness and gives rise to a yield stress. 

To model the response of the fiber, we propose to decouple the overall
behavior of the spider silk fiber into two networks that deform in
parallel: (1) an elasto-plastic network of intermolecular and intramolecular
bonds and (2) an elastic network of polypeptide chains that deform
entropically. In this context, the bonds can be viewed as frictional
elements that dissipate energy and lead to the accumulation of plastic
strain during the initial loading of the fiber, whereas the chains
enable large deformations to occur once mobility is gained.

One of the key findings of this contribution is the delineation of
the mechanisms that govern the relaxation and recovery mechanisms
in the silk fiber. We demonstrate that the \textquotedbl healing\textquotedbl{}
of the fiber in a traction free state stems from two factors: (1)
the reformation of intermolecular and intramolecular bonds and (2)
a microstructural reorganization of the chains. During this process,
the intermolecular bonds fix the microstructure and establish a new
stable equilibrium configuration. The reorganization of the chains
and the reformation of the bonds explain the increase in stiffness
and yield stress in subsequent loadings. 

To demonstrate the merit of this work, we follow experimental findings
and propose explicit expressions for the stiffness (Young's modulus)
and the yield stress as a function of the relaxation stretch. Next,
these expressions and the model predictions for the response of fibers
under continuous loading and cyclic uniaxial extension are compared
to experimental data on silk fibers from \emph{Argiope bruennichi}.
The model is capable of capturing the continuous load to failure and
the response of the fiber under multiple cycles. It is also worth
noting that the model can be readily amended to capture the dependency
on strain rate through the elasto-plastic network.

In conclusion, the main contributions of this work are three-fold:
(1) providing a physical explanation for the origin of hysteresis
and residual strain in spider silk fibers, linking them to microstructural
evolution and rearrangement, (2) quantifying the recovery capabilities
of the network, revealing how bond reformation restores and enhances
mechanical properties, and (3) paving the way towards a robust predictive
tool for the design of bio-inspired synthetic fibers that require
tunable mechanical properties such as stiffness, yield stress, strength,
energy dissipation, and recovery.

\paragraph{Acknowledgments}

JPR was funded by Ministerio de Ciencia e Innovaci�n (Spain) (grant
PID2023-152058OB-I00) and by the European Union\textquoteright s EIC-Pathfinder
Programme under the project THOR (Grant Agreement number 101099719). 

\appendix

\section{Integration from the chain to the network level \label{sec:micro-sphere} }

To bridge the scales between the local response of individual polypeptide
chains and the macroscopic behavior of the fiber, we employ the micro-sphere
technique \citep{baza&oh86ZAMM}. This numerical approach facilitates
the integration of chain-level quantities over the unit sphere to
determine the overall network response. Specifically, rather than
integrating a given quantity over all possible spatial directions,
this method provides an approximation by summing over a discrete set
of $m$ orientation vectors $\dirTref^{\left(i\right)}$ with associated
weighting factors $w^{(i)}$. Accordingly, the average of a microscopic
quantity $\bullet$ can be computed via 

\begin{equation}
\langle\bullet\rangle=\frac{1}{4\pi}\int_{A}\bullet\,\d A\approx\sum_{i=1}^{m}\bullet^{(i)}w^{(i)},
\end{equation}
where the weights are constrained by $\sum_{i=1}^{m}w^{(i)}=1$. In
isotropic networks, we require that $\left\langle \dirTref^{\left(i\right)}\right\rangle =\mathbf{0}$
and $\left\langle \dirTref^{\left(i\right)}\otimes\dirTref^{\left(i\right)}\right\rangle =1/3\mathbf{I}$. 

In the following, the micro-sphere technique is used to determine
the stress in the chain networks (see Eq. \ref{eq:total_stress_chain_network}).
\citet{baza&oh86ZAMM} showed that a specific choice of $m=42$ orientation
vectors provides sufficient accuracy, and we adopt that conclusion
in this work. The representative directions are listed in Table 1
of \citet{baza&oh86ZAMM}.

\bibliographystyle{biochem}
\bibliography{SC_bib}

\end{document}